\documentclass[aps,showpacs,twocolumn,superscriptaddress,prl]{revtex4}
\usepackage{graphicx,color,epsfig,rotate}
\usepackage{amssymb,amsmath,bm}

\begin{document}

\title{Lifetime of Gapped Excitations in Collinear Quantum Antiferromagnet}

\author{A. L. Chernyshev}
\affiliation{Department of Physics, University of California, Irvine, California
92697, USA}
\affiliation{Service de Physique Statistique, Magn\'etisme et Supraconductivit\'e,
UMR-E9001 CEA-INAC/UJF, 17 rue des Martyrs, 38054 Grenoble Cedex 9, France}
\author{M. E. Zhitomirsky}
\affiliation{Service de Physique Statistique, Magn\'etisme et Supraconductivit\'e,
UMR-E9001 CEA-INAC/UJF, 17 rue des Martyrs, 38054 Grenoble Cedex 9, France}
\author{N. Martin}
\affiliation{Service de Physique Statistique, Magn\'etisme et Supraconductivit\'e,
UMR-E9001 CEA-INAC/UJF, 17 rue des Martyrs, 38054 Grenoble Cedex 9, France}
\author{L.-P. Regnault}
\affiliation{Service de Physique Statistique, Magn\'etisme et Supraconductivit\'e,
UMR-E9001 CEA-INAC/UJF, 17 rue des Martyrs, 38054 Grenoble Cedex 9, France}
\date{\today}

\begin{abstract}
We demonstrate that local modulations of magnetic couplings have a profound effect on the temperature
dependence of the relaxation rate of optical magnons in a wide class of antiferromagnets
in which gapped excitations coexist with acoustic spin waves.
In a two-dimensional collinear antiferromagnet
with an easy-plane anisotropy, the disorder-induced relaxation rate of the gapped mode,
$\Gamma_{\rm imp}\approx\Gamma_0+A\left(T\ln T\right)^2$,
greatly exceeds the magnon-magnon damping, $\Gamma_{\rm m-m}\approx BT^5$,  negligible at low temperatures.
We measure the lifetime of gapped magnons in a prototype $XY$ antiferromagnet
BaNi$_2$(PO$_4$)$_2$ using  a high-resolution neutron-resonance spin-echo technique and find
experimental data in close accord with the theoretical prediction.
Similarly strong effects of disorder in the three-dimensional case and in noncollinear
antiferromagnets are discussed.
\end{abstract}
\maketitle

{\it Introduction.}---%
The recent development of the neutron-resonance spin-echo technique has led
to dramatic improvement of the energy resolution in neutron-scattering
experiments \cite{Bayrakci06,Keller06,Haug10,Nafradi11}.
When applied to elementary excitations in magnetic insulators, this technique
allows one to measure magnon linewidth with the $\mu$eV accuracy compared
to the meV resolution of a typical triple-axis spectrometer.
Damping of quasiparticles depends fundamentally on the strength of their interactions with each
other and with impurities, information not accessible directly by other measurements.
Although theoretical studies of magnon damping in antiferromagnets (AFs) go back to
the 1970s \cite{HKHH,Rezende}, a comprehensive comparison between theory
and experiment is still missing, mainly due to the lack of experimental data.

Magnon-magnon scattering is traditionally
viewed as the leading source of temperature-dependent
magnon relaxation rates in AFs \cite{HKHH,Rezende}.
Another common relaxation mechanism
in solids is the lattice disorder, which is responsible for a variety of the low-temperature effects,
such as  residual resistivity of metals \cite{Bass} and finite linewidth of
antiferromagnetic resonances \cite{WFW}.
However, {\it  temperature-dependent} effects
of disorder are usually neglected because of the higher powers
of $T$ in   impurity-induced relaxation rates
compared to  leading scattering mechanisms
and  of the presumed dilute concentration and weakness of  disorder.
The closest analogy is the resistivity of metals, in which the $T=0$ term is due to
lattice imperfections and the temperature-dependent part is due to  quasiparticle scattering.

\begin{figure}[b]
\includegraphics[width=0.3 \columnwidth]{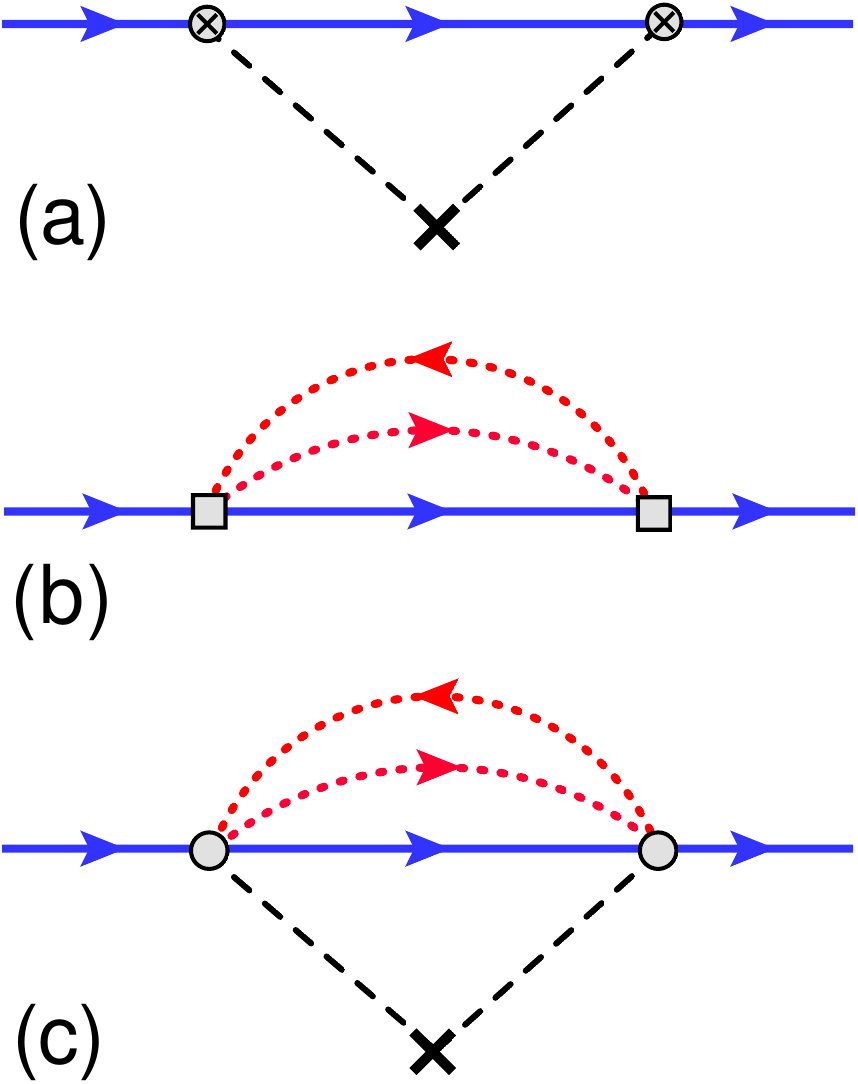}\hskip 0.9cm
\includegraphics[width=0.38\columnwidth]{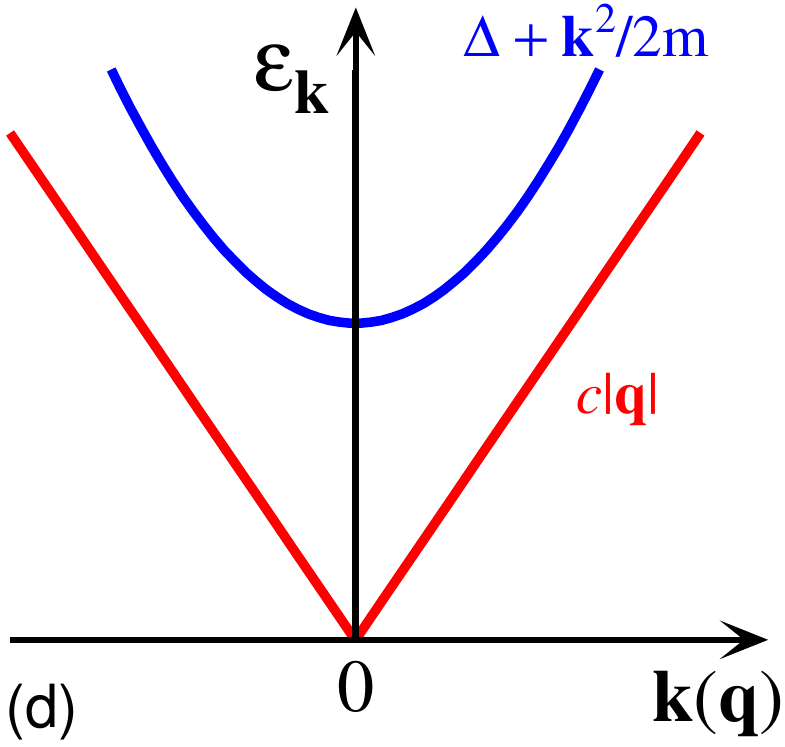}
\caption{(Color online) (a)-(c) Diagrams representing impurity, magnon-magnon,
and impurity-assisted 
scattering of the optical magnon (solid lines). 
Dotted lines are acoustic magnons. (d) Schematic energy spectrum of
the model (\ref{H}).}
\label{fig:diagrams}
\end{figure}

In this work, we demonstrate that scattering on the spatial modulations of magnetic couplings should completely
dominate the low-temperature relaxation rate of gapped excitations in a wide class of AFs.
 Such modulations,
produced by random lattice distortions, yield  scattering potential for propagating
magnons and, at the same time,  modify locally their interactions. For an illustration,
we consider an example of the two-dimensional (2D) easy-plane AF with one
acoustic and one gapped excitation branch. In addition to   potential scattering,
responsible for a finite damping $\Gamma_0\propto n_i$ of optical magnons,
see Fig.~\ref{fig:diagrams}(a), there exists
an impurity-assisted temperature-dependent scattering of gapped magnons on
thermally-excited acoustic spin waves, see Fig.~\ref{fig:diagrams}(c), which yields
$\Gamma_{\rm imp}(T)\!\propto \! n_i T^2\ln^2 T$. Despite the presumed smallness
of impurity concentration $n_i$, at low temperatures this
mechanism dominates over the conventional magnon-magnon scattering, Fig.~\ref{fig:diagrams}(b),
which carries a much higher power of temperature: $\Gamma_{\rm mm}\!\propto\!T^5$.
We have   performed resonant neutron spin-echo measurements with a few $\mu$eV resolution
on a high-quality sample of BaNi$_2$(PO$_4$)$_2$, a prototype 2D planar
AF \cite{Regnault}.
We find that the theory describes very well the experimental data
for the linewidth of optical magnons.
Similar dominance of the impurity-assisted magnon-magnon
scattering should persist in the 3D AFs and is even more pronounced in the
noncollinear AFs.
We propose further experimental tests of this mechanism.

{\it Theory.}---%
We begin with the spin Hamiltonian of a collinear AF with an easy-plane
anisotropy induced by the single-ion term $D>0$:
\begin{eqnarray}
\label{H}
{\cal H} = \sum_{\langle ij\rangle} J_{ij}\,{\bf S}_i\cdot {\bf S}_j
         +  D \sum_i \left(S^z\right)^2\, .
\end{eqnarray}
 Two examples are the
nearest-neighbor AFs on square and  honeycomb lattices.
The latter model, with the non-frustrating third-neighbor
exchange,  is relevant to the spin-1 antiferromagnet
$\rm BaNi_2(PO_4)_2$ \cite{Regnault} discussed below.

As a consequence of broken $XY$ symmetry, excitation spectrum in the ordered
antiferromagnetic state possesses acoustic ($\alpha$) and gapped ($\beta$) magnon branches:
\begin{equation}
\varepsilon^\alpha_{\bf k}\approx c|{\bf k}|\ , \qquad
\varepsilon^\beta_{\bf k}\approx \Delta+\frac{{\bf k}^2}{2m}\ ,
\label{wk}
\end{equation}
see Fig.~\ref{fig:diagrams}(d) for a sketch. Explicit expressions for $c$, $\Delta$, and $m$ for
BaNi$_2$(PO$_4$)$_2$  are provided in \cite{suppl}. 

Defects are present in all  crystals. While vacancies and substitutions may be 
eliminated in some materials, inhomogeneous lattice distortions remain
an intrinsic source of disorder, inducing
weak random variations $\delta J$ and $\delta D$ of microscopic parameters
in the spin Hamiltonian (\ref{H}) \cite{Kohama11}. Both types of randomness
have qualitatively the same effect on magnon lifetimes.
For example, local modification of the single-ion anisotropy
$\delta D(S^z_\ell)^2$ generates scattering potential for magnons
\begin{eqnarray}
\label{Himp1}
{\cal H}_2^{\rm imp}=\sum_{\bf k,k'} e^{i ({\bf k-k'}){\bf R}_\ell}\,
U_{{\bf k}{\bf k'}}\, c_{\bf k'}^\dag c^{_{}}_{\bf k} \ ,
\end{eqnarray}
where $c_{\bf k}\! = \!\alpha_{\bf k}(\beta_{\bf k})$,
$U_{{\bf k}{\bf k'}}\! =\! \delta DS\,(u_{\bf k}\!+\!v_{\bf k})(u_{\bf k'}\!+\!v_{\bf k'})$, and
$u_{\bf k},v_{\bf k}$ are the Bogolyubov transformation parameters.
For optical magnons at ${\bf k},{\bf k}'\to 0$, the momentum dependence is not important,
$U_{{\bf k}{\bf k'}}\!=\!O(\delta D)$.
For bond disorder, all
expressions are the same with a substitution $\delta D\!\to\!\delta J$
and an additional phase factor, which depends on bond orientation
and disappears after impurity averaging.

For the gapped magnons with ${\bf k}\rightarrow 0$, scattering amplitude in the second Born
approximation, Fig.~\ref{fig:diagrams}(a), averaged over spatial distribution of impurities is \cite{Mahan}
\begin{eqnarray}
\label{tau_0}
\Gamma^{\rm imp}_{\bf k}\approx \Gamma_0\propto  n_i
\overline{U}^2_i\,
\frac{m \,\omega_{\rm max}^2}{\Delta^2}\, ,
\end{eqnarray}
where $n_i$ is the impurity concentration,
$\overline{U}_i=O(\delta J,\delta D)$ is the averaged impurity potential,
and $\omega_{\rm max}$ is the magnon bandwidth \cite{suppl}.
Thus, in 2D, conventional impurity scattering results in a finite zero-temperature
relaxation rate of the gapped magnons.

At low temperatures, the principal scattering channel for optical magnons
is due to collisions with the thermally excited acoustic spin waves with
$cq \sim T\ll \Delta$. All other processes are either forbidden
kinematically or exponentially suppressed. In this case we can consider
only $\beta\alpha\rightarrow\beta\alpha$ terms in the magnon-magnon interaction:
\begin{eqnarray}
\label{Hmm}
&&{\cal H}_4^{\rm mm}=\sum_{\bf k+q=k'+q'} V^{\rm mm}_{\bf kq;k'q'}
\beta_{\bf k'}^\dag\alpha_{\bf q'}^\dag \alpha^{\phantom \dag}_{\bf q}
\beta^{\phantom \dag}_{\bf k} \, , \\
&&\label{1H4imp}
{\cal H}_4^{\rm imp}=\sum_{\bf kq,k'q'} e^{i\Delta{\bf k}{\bf R}_\ell}\, V^{\rm\, imp}_{\bf k,q;k',q'}
\beta_{\bf k'}^\dag\alpha_{\bf q'}^\dag \alpha^{\phantom \dag}_{\bf q}
\beta^{\phantom \dag}_{\bf k} \, , \ \ \ \
\end{eqnarray}
where the first and the second row  correspond to the conventional  and to the 
impurity-assisted magnon-magnon scattering, respectively, with
$\Delta{\bf k}\!=\!{\bf k}\!+{\bf q}\!-{\bf q'}\!-{\bf k'}$. 
The latter is of the {\it same} origin as the conventional impurity scattering in  (\ref{Himp1}) 
since $\delta D$ and $\delta J$ also modify locally interactions among magnons \cite{suppl}.
In the one-loop approximation,   (\ref{Hmm}) and (\ref{1H4imp})
yield the self-energies of Figs.~\ref{fig:diagrams}(b) and (c). Applying
standard Matsubara technique, relaxation rates can be expressed as
\begin{eqnarray}
\label{1Gmm}
&&\Gamma^{{\rm mm}}_{\bf k} = \pi \sum_{\bf qq'}
\left|V^{\rm mm}_{\bf kq;k'q'}\right|^2 N^{\bf q}_{{\bf k'}{\bf q'}}\, \delta(\Delta\varepsilon)\ ,\\
\label{1Gimp}
&&\Gamma^{{\rm imp},T}_{\bf k} = \pi n_{i} \sum_{\bf qq'k'}
\left|\bar{V}^{\rm\, imp}_{\bf kq;k'q'}\right|^2
N^{\bf q}_{{\bf k'}{\bf q'}} \, \delta(\Delta\varepsilon) \ ,
\end{eqnarray}
where $\Delta\varepsilon = \varepsilon_{\bf k}\!+\varepsilon_{\bf q}\! -\varepsilon_{\bf q'}\!-\varepsilon_{\bf k'}$,
$N^{\bf q}_{{\bf k'}{\bf q'}}= n_{\bf q}(1+n_{\bf q'}\!+n_{\bf k'})-n_{\bf q'} n_{\bf k'}$,
and $n_{\bf q}$ is the Bose factor. 

There are two important differences between $\Gamma^{{\rm mm}}$ and
$\Gamma^{{\rm imp},T}$ in (\ref{1Gmm}) and (\ref{1Gimp}). First,
the total momentum is not conserved for impurity scattering. This relaxes
kinematic constraints of the 4-magnon scattering processes, but requires
instead integration over the extra independent  momentum $\bf k'$.
Second and most crucial, interaction vertices $V^{\rm mm}_{\bf kq;k'q'}$ and
$V^{\rm imp}_{\bf kq;k'q'}$ show very different long-wavelength behavior as
${\bf q}$, ${\bf q}'\rightarrow 0$.
We calculate them using the approach
similar to \cite{HKHH,Rezende}, and find that in the long-wavelength limit
magnon-magnon interaction (\ref{Hmm}) is
$V^{\rm mm}_{\bf kq;k'q'}\propto \sqrt{qq'}$, in accordance with
the hydrodynamic limit \cite{LLIX}.
However, for the impurity-assisted scattering (\ref{1H4imp}),  interaction is
$V^{\rm\, imp}_{\bf kq;k'q'}\propto 1/\sqrt{qq'}$. This can be understood as
a consequence of an effective long-range  potential for acoustic magnons
produced by the gaped magnon while in the vicinity of an impurity.

The leading $T$-dependence of $\Gamma^{{\rm mm}}_{{\bf k}\rightarrow 0}$ and
$\Gamma^{{\rm\, imp},T}_{{\bf k}\rightarrow 0}$ can be calculated now
using (\ref{wk}) and approximating  interaction vertices with
their long-wavelength expressions. The main contribution to the integrals in
(\ref{1Gmm}) and (\ref{1Gimp}) is determined by acoustic magnons with
$q, q'\! \sim\! T/c$. Then, a straightforward power counting  yields
\begin{equation}
\label{1Gmm_est1}
\Gamma^{{\rm mm}}_{{\bf k}\rightarrow 0}\approx B\left(\frac{T}{\omega_{max}}\right)^5,
\end{equation}
where $B\sim \omega_{max}$ \cite{suppl}. Thus, the inverse lifetime of an optical magnon
is proportional to $T^5$ in 2D. A generalization to higher dimensions gives
$\Gamma^{{\rm mm}}\propto T^{2D+1}$. The $T^7$-law for the relaxation rate
of optical magnons in 3D AFs was previously predicted in
\cite{Baryakhtar73}. We note that for a given model, the effect of magnon-magnon scattering  in (\ref{1Gmm_est1})
can be calculated using microscopic parameters, thus putting strict bounds on its
magnitude.

The same calculation for  $\Gamma^{{\rm imp},T}_{{\bf k}\rightarrow 0}$ proceeds via
the following integral:
\begin{eqnarray}
\label{1Gimp_est}
\Gamma^{{\rm imp},T}_{{\bf k}\rightarrow 0}\approx
\frac{n_i\overline{U}_i^2}{8\pi^2}
\int_q\int_{q'} n_{\bf q}\left(n_{{\bf q}'}+1\right) \int_0^{\infty} k'\, dk'   \,
\delta\left(\Delta \varepsilon\right), \
\end{eqnarray}
where $\int_q\!=\!\int_0^\Lambda dq$ with $\Lambda\!\sim\!\pi/a$,
$\Delta \varepsilon\!=\!cq\!-\!cq' \!-\! k'^2/2m$, and we used
the relation between  $\bar{V}^{\rm\, imp}_{\bf kq;k'q'}$ in (\ref{1H4imp}) and $U_{{\bf k}{\bf k'}}$
in (\ref{Himp1}).
The na\"{\i}ve power counting in (\ref{1Gimp_est}) already gives $\Gamma^{{\rm imp},T}\!\propto\!T^2$,
while a more careful consideration shows further enhancement of the scattering as
the integrals formally diverge [logarithmically] in the
${q}\rightarrow 0$ region, demonstrating an important role of the  long-wavelength magnons  in 2D.
This divergence is similar to the one in the problem of finite $T_N$ ordering temperature in 2D and is
regularized similarly by  introducing  low-energy cutoff.
The cutoff is either due to a 3D-crossover  as in the case of some cuprates \cite{cuprates}, or  a
 weak  in-plane  anisotropy that induces small gap   $\omega_0$ in the acoustic branch,  the case
directly relevant to the current work \cite{Regnault,Ikeda}.

Combining  (\ref{tau_0}) and (\ref{1Gimp_est}) we obtain  impurity-induced
 relaxation rate of gapped magnons
\begin{eqnarray}
\label{1Gimp_tot}
\Gamma^{{\rm imp}}\approx \Gamma_0 + A\biggl(\frac{T}{\omega_{max}}\biggr)^2
\left[\biggl(\ln{\frac{T}{\omega_0}}\biggr)^2 +\frac{\pi^2}{3}\right]
\, ,
\end{eqnarray}
where both $\Gamma_0$ and $A$ are proportional to $n_i$ and to the average
strength of disorder $\overline{U}_i^2$.
As a result, the impurity scattering leads to a relaxation rate that carries a significantly lower power of
temperature than the magnon-magnon scattering mechanism. Therefore, despite
possible  smallness of the combined impurity concentration and strength, it should dominate not only the $T=0$
lifetime of the gapped magnon, but also its temperature dependence in the entire low-temperature regime.
A qualitative prediction of our consideration is that $\Gamma_0$ and $A$ in (\ref{1Gimp_est})
should be of the same order since both terms are
related to disorder. In addition, for samples of the same material of different quality,
they must scale with the amount of structural disorder in a correlated way.

In the 3D case, impurity-assisted mechanism (\ref{1Gimp_est}) gives
$\Gamma^{{\rm imp},T}_{3D}\propto T^{9/2}$, still 
dominating the 3D magnon-magnon relaxation rate $\Gamma^{{\rm mm}}_{3D}\propto T^7$ discussed above.

{\it Experiment.}---%
The experimental part of our work is devoted to the neutron spin-echo
measurements of the magnon lifetime in $\rm BaNi_2(PO_4)_2$.
This material is a layered quasi-2D AF with a honeycomb lattice
of spin-1 Ni$^{2+}$ ions and  N\'eel temperature $T_N\!\approx \!25$~K.
A comprehensive review of the physical properties of $\rm BaNi_2(PO_4)_2$ is presented in \cite{Regnault}.
Its excitation spectrum
has an optical branch with the gap $\Delta \approx 32$~K and an acoustic
mode, as is sketched in Fig.~\ref{fig:diagrams}(d).
The fit of the magnon dispersion yields
the following microscopic parameters:
$J_1\!=\!0.38$~meV and  $J_3\!=\!1.52$~meV are exchanges between first- and third-neighbor spins,
and  $D\!=\!0.32$~meV is the single-ion anisotropy.
The thermodynamic properties  of BaNi$_2$(PO$_4$)$_2$ follow the 2D behavior down
to $T\!\alt\! 1$K and a small gap in the acoustic branch, $\omega_0\!\approx\! 2$K,
due to weak in-plane anisotropy is consistent with
the value of the ordering temperature \cite{Regnault}.

\begin{figure}[b]
\includegraphics[width=0.9\columnwidth]{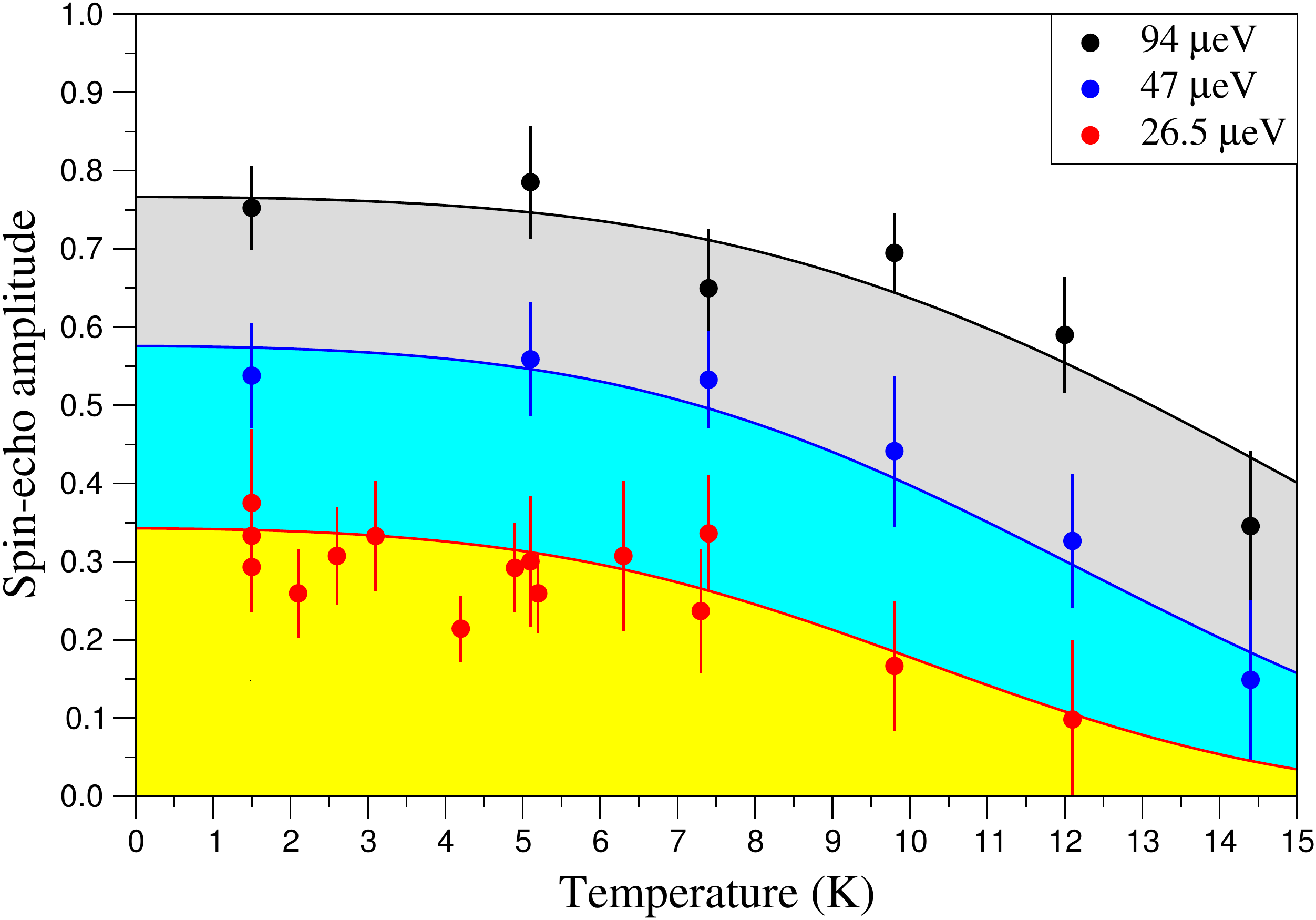}
\caption{(Color online) Temperature dependence of the polarization (spin-echo amplitude) of the
neutron beam  $P(T)$ for several representative spin-echo energies. }
\label{fig:Polarization}
\end{figure}

The spin-echo experiments were performed on the triple-axis spectrometer IN22 (ILL, Grenoble)
by using ZETA neutron resonance  spin-echo option \cite{Martin11}. The incident neutron
beam was polarized and the scattered beam analyzed from (111) reflection of $\rm Cu_2MnAl$
Heusler alloy focusing devices. We used a fixed-$k_f$ configuration, with $k_f=2.662$~\AA$^{-1}$
or $k_f=1.97$~\AA$^{-1}$. Different rf-flipper configurations were used in order to adapt
the spin-echo time (energy)  $t_{\rm NSE}$ ($\varepsilon_{\rm NSE}=h/t_{\rm NSE}$)
to the magnetic excitation lifetimes, typically in the range of $5-50\,$ps ($130-13\,\mu$eV).
As for any spin-echo experiment \cite{Mezei72,Golub87},
the measurement of  the neutron polarization (spin-echo amplitude) after the scattering, $P(t_{\rm NSE})$,
provides us with a direct access to the correlation function
$S({\bf q},t_{\rm NSE})$. For a spin-wave excitation described by a Lorentzian function
in energy of half width $\Gamma$, one can show that $P(\varepsilon_{\rm NSE})\! =\!
P_0(\varepsilon_{\rm NSE}) \exp({-\Gamma/\varepsilon_{\rm NSE}})$, in which the prefactor
$P_0$ depends on the spin-echo resolution.

\begin{figure}[b]
\includegraphics[width=0.99\columnwidth]{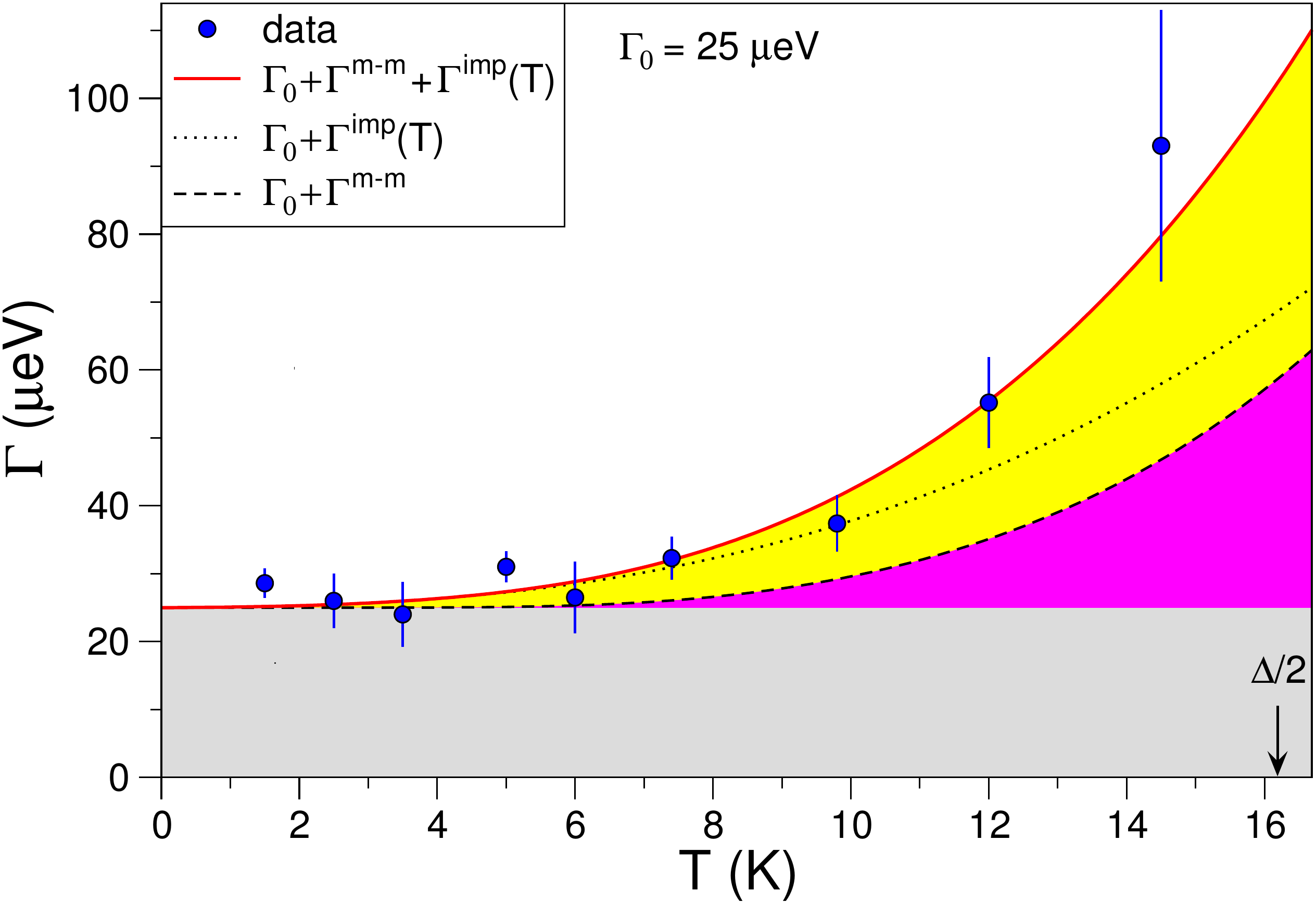}
\caption{(Color online) Temperature dependence of the relaxation rate $\Gamma$
of the optical magnon with ${\bf k}\approx 0$ in BaNi$_2$(PO$_4$)$_2$.
Full line is the best theoretical fit including all contributions with parameters
described  in the text.
Dashed and dotted lines indicate separate contributions of magnon-magnon
and impurity-assisted magnon-magnon scattering. }
\label{fig:Gamma}
\end{figure}

For our measurements, we have used a $2\;\textrm{cm}^3$ single crystal of $\rm BaNi_2(PO_4)_2$ oriented with
the ${\bf a}^*$ and ${\bf c}^*$ reciprocal axes in the scattering plane. The spin-echo data were taken at
the antiferromagnetic scattering vector ${\bf Q}_{\rm AF}=(1,0,0)$ and the energy transfer
$\Delta E=3$~meV corresponding to the bottom of the dispersion curve of the gapped mode \cite{Regnault}.
In determining   the spin-echo amplitudes, neutron intensities were corrected for
the inelastic background, measured at the scattering vector ${\bf Q}_{\rm AF}$  and the energy
transfer $\Delta E=5$~meV. Results of the temperature dependence of spin-echo amplitudes
for  several representative $\varepsilon_{\rm NSE}$'s are shown in Fig.~\ref{fig:Polarization}.
Solid lines are the fits of the spin-echo amplitudes with $P\! =\!P_0 
e^{-\Gamma/\varepsilon_{NSE}}$ using relaxation rate 
in the functional form given by (\ref{1Gmm_est1}) and (\ref{1Gimp_tot}), 
$\Gamma\!=\!\Gamma^{\rm mm}+\Gamma^{\rm imp}$, which we
discuss next. Using the full set of $P(T,\varepsilon_{\rm NSE})$ data, experimental results for $\Gamma(T)$
are extracted from the fits of $\ln (P)$ vs $\varepsilon_{\rm NSE}$ at fixed temperatures. 
These results are presented in our Fig.~\ref{fig:Gamma}  together  with the theoretical fits.

{\it Comparison.}---%
The relaxation rate approaches  the constant value of $\Gamma_0\!\approx\!25$ $\mu$eV
at $T\!\rightarrow\! 0$, in agreement with the expectation  (\ref{tau_0}) for the gapped mode in 2D.
The low-$T$ dependence of the relaxation rate is following the power law much slower than $T^5$. The quality of   
the free-parameter fit of $\Delta\Gamma\!=\!\Gamma(T)\!-\!\Gamma_0$ with just the 
$T^5$ law is not satisfactory for either $\Gamma(T)$ or $P(T)$'s in Figs.~\ref{fig:Gamma}  
and \ref{fig:Polarization}, and the magnitude of  $\Delta\Gamma$
also requires an unphysically large values 
of the magnon-magnon scattering parameter $B$ in (\ref{1Gmm_est1}), exceeding theoretical estimates roughly tenfold.
On the other hand, $T^2\ln^2T$ law gives much more satisfactory fits in the low- and intermediate-$T$ regime  up to 12 K
in both $\Gamma(T)$ and $P(T)$, shown as a separate fit by the dotted line in Fig.~\ref{fig:Gamma}.
The best fit of $\Gamma(T)$, given by solid line, is the sum of the 
magnon-magnon and impurity-scattering effects from (\ref{1Gmm_est1}) and (\ref{1Gimp_tot}), with the
magnon-magnon and impurity-assisted parameters $B\!=\!15$ meV and $A\!=\!90$ $\mu$eV, respectively.
The same $\Gamma(T)$ is used in all three curves of $P(T)$ in Fig.~\ref{fig:Polarization}, the original data from which 
experimental $\Gamma(T)$ is extracted. Magnon bandwidth $\omega_{max}\!=\!64$ K and the low-energy cutoff 
 $\omega_0\!=\!2$ K, equal to the gap in the acoustic branch, were used.

Two remarks are in order concerning the role of the magnon-magnon relaxation rate used in Fig.~\ref{fig:Gamma}. 
First,  fits of $\Gamma(T)$ in Fig.~\ref{fig:Gamma}
also include a contribution from scattering off the thermally excited optical magnons, which is
given by $\Gamma^{\rm rr}\!=\!C\left(\frac{T}{\Delta}\right)e^{-\Delta/T}$ \cite{suppl}. Its contribution is roughly 
equal to that of the $T^5$-term (\ref{1Gmm_est1}) at $T\!=\!16$K ($=\!\Delta/2$), but diminishes faster at lower $T$.
In the fit of $\Gamma(T)$ we use the value of $C\!=\!260$ $\mu$eV,  about three times  the theory 
estimate: $C^{th}\!\approx\!70$ $\mu$eV. 
Second, the theoretical estimate of  the magnon-magnon interaction parameter in $T^5$ law (\ref{1Gmm_est1}) is 
$B^{th}\!\approx\!6$ meV, again factor 2.5 smaller than the one used in the fit ($B\!=\!15$ meV).
  Altogether, the magnon-magnon contribution to $\Gamma(T)$, 
shown by the dashed line and the corresponding color shading in Fig.~\ref{fig:Gamma}, is likely a generous
overestimate of its actual role in the relaxation.

Still, the contribution of the impurity-assisted mechanism in $\Gamma(T)$ is very strongly pronounced 
and is not explicable by the conventional scattering mechanisms. For example, at 12K the impurity 
scattering accounts for at least 2/3 of the temperature-dependent part of $\Gamma(T)$. The parameter
of the impurity-assisted term in (\ref{1Gimp_tot}) used in the fit is 
$A\!=\!90\mu$eV, which is of the same order with the constant impurity term $\Gamma_0$, meeting our expectations 
outlined above. This is, again, the strong argument that both the constant and the
$T$-dependent terms in the relaxation rate must have the same origin, giving
further support to the consistency of our explanation of the data. 

The values of $A$ and $\Gamma_0$ cannot be determined theoretically as the impurity concentration and
strength are, generally, unknown. However, another consistency check is possible:
the ratio of $\Gamma_0$ to a characteristic energy scale of the problem, $\omega_{max}$, should give,
according to (\ref{tau_0}),  an estimate of the cumulative measure of disorder concentration and its
strength: $n_{\rm imp}  (\overline{\delta D}/D)^2\!\approx\!\Gamma_0/\omega_{max}\!\approx\! 5\cdot 10^{-3}$.
This translates into a  reasonable estimate of the disorder and its strength in BaNi$_2$(PO$_4$)$_2$: 
modulation of magnetic couplings is equivalent to 
half of a percent of sites having $\delta D$ ($\delta J$) of order $D$ ($J$). 
The amount of structural distortion in BaNi$_2$(PO$_4$)$_2$ \cite{Larmor12} is consistent
with the  magnitude of such variations of magnetic couplings, given the
strong spin-lattice coupling
in this material.

{\it Other systems.}---%
We propose that similar, and even stronger, effects of disorder in the relaxation rate must be present
in the 2D noncollinear AFs, in which magnon-magnon interactions  acquire
the so-called cubic interaction terms \cite{ZhCh},
absent in the  collinear AFs considered above.
The self-energies associated with such interaction are the same as in Figs.~\ref{fig:diagrams}(b) and (c),
but with two intermediate lines instead of three.
With the long-wavelength behavior of the impurity interaction to follow
$\delta V_3({\bf k},{\bf q})\!\propto\! 1/\sqrt{q}$, as in the considered case, a qualitative consideration similar to
(\ref{1Gimp_est}) leads to:
\begin{eqnarray}
\label{G3imp}
\Gamma^{{\rm imp},T}_{{\bf k}\rightarrow 0}\approx A_3\,
 \left(\frac{T}{\omega_{max}}\right)\,\ln{\frac{T}{\omega_0}}\, ,
\end{eqnarray} 
where $A_3\!\propto\! n_{\rm imp}(\overline{\delta D}/D)^2$,
an even lower power of $T$.
Since the canting of spins can be induced by the external field, we propose an experimental investigation 
of the effect of such a field on the relaxation rate. For the 3D noncollinear AFs we predict
$\Gamma^{{\rm imp},T}\propto T^{5/2}$.

Recent neutron spin-echo experiment in a Heisenberg-like AF MnF$_2$ \cite{Bayrakci06} have
reported significant discrepancies between  measured relaxation rates and predictions of the
magnon-magnon scattering theory \cite{HKHH,Rezende}, precisely in the regime of low-$T$ and
small-${\bf k}$ where the theory is assumed to be most reliable.
Although   the current work concerns the dynamics of strongly gapped
excitations and our results are not directly transferable to the case of MnF$_2$, we have, nevertheless,
presented a general case in which the magnon-magnon scattering mechanism is completely overshadowed
by  impurity scattering, thus suggesting a similar consideration in  other systems.

{\it Conclusions.}---%
To conclude, we have presented strong evidence of the general situation in which temperature-dependence of the
relaxation rate of a magnetic excitation is completely dominated by the effects induced by  simple structural disorder. Our results
are strongly supported by the available experimental data. Further theoretical and experimental studies are suggested.

This work was initiated at the Max-Planck Institute for the Physics of Complex Systems during the activities of the Advanced Study
Group Program on ``Unconventional Magnetism in High Fields,'' which we would like to thank for hospitality.
The work of A.~L.~C. was supported by the DOE under Grant No. DE-FG02-04ER46174.



\newpage
\onecolumngrid
\begin{center}
{\large\bf Lifetime of Gapped Excitations in Collinear Quantum Antiferromagnet:\\
Supplemental Information}\\ 
\vskip0.35cm
A. L. Chernyshev$^{1,2}$, M. E. Zhitomirsky$^2$, N. Martin$^2$, and L.-P. Regnault$^2$\\
\vskip0.15cm
{\it \small $^1$Department of Physics, University of California, Irvine, California
92697, USA\\
$^2$Service de Physique Statistique, Magn\'etisme et Supraconductivit\'e,\\
UMR-E9001 CEA-INAC/UJF, 17 rue des Martyrs, 38054 Grenoble Cedex 9, France}\\
{\small (Dated: June 20, 2012)}\\
\vskip -0.1cm \
\end{center}
\twocolumngrid

\section*{Spin Hamiltonian}

Here we briefly outline basic steps and main results of the spin-wave calculations
for the energy spectrum and the magnon relaxation rates of 
the $J_1$--$J_3$ antiferromagnet on a honeycomb lattice. The harmonic spin-wave
analysis of the nearest-neighbor
Heisenberg honeycomb-lattice antiferromagnet can be found, for example, in
\cite{Weihong91sup}.

\begin{figure}[b]
\centerline{
\includegraphics[width=0.8\columnwidth]{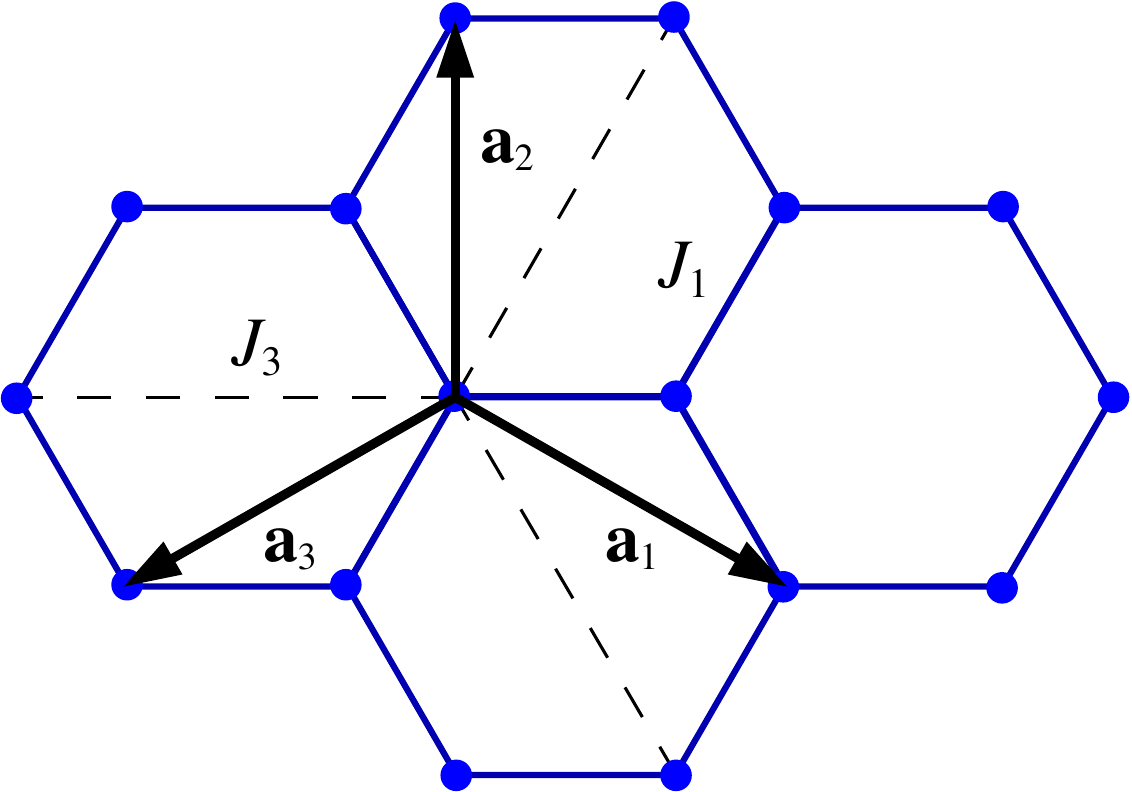}}
\caption{$J_1$--$J_3$ model in a honeycomb lattice.}
\label{suppl:lattice}
\end{figure}

Geometry of exchange bonds of the considered model is schematically shown in
Fig.~\ref{suppl:lattice}. The unit cell of the antiferromagnetic structure
coincides with the crystal unit cell and contains two oppositely aligned spins
${\bf S}_{1,i}$ and ${\bf S}_{2,i}$ in positions  $(0,0)$ and
$\bm{\rho} = (a/\sqrt{3},0)$. The elementary translation vectors are defined
as ${\bf a}_1= a(\sqrt{3}/2,-1/2)$ and ${\bf a}_2= a(0,1)$.
The lattice constant in BaNi$_2$(PO$_4$)$_2$ is equal to $a = 4.81$~\AA.
The reciprocal lattice basis is ${\bf b}_1= 4\pi/(\sqrt{3}a)(1,0)$
and ${\bf b}_2= 2\pi/(\sqrt{3}a)(1,\sqrt{3})$. The volume of the Brilouin zone
is $V_{\rm BZ} = 8\pi^2/\sqrt{3}a^2$.

The spin Hamiltonian includes Heisenberg exchange interactions between first-
and third-neighbor spins together with the single-ion anisotropy:
\begin{eqnarray}
\hat{\cal H} & = & J_1 \sum_i {\bf S}_{1,i}\cdot ({\bf S}_{2,i} +
{\bf S}_{2,i-1}+{\bf S}_{2,i+3})
\nonumber \\
& + & J_3 \sum_i {\bf S}_{1,i}\cdot ({\bf S}_{2,i+2} + {\bf S}_{2,i-2} +
{\bf S}_{2,i-1+3})
\label{suppl:H} \\
& + & D \sum_i \bigl[(S_{1,i}^z)^2 +(S_{2,i}^z)^2   \bigr] \ .
\nonumber
\end{eqnarray}
Here ${\bf S}_{2,i-1}$ denotes spin in the unit cell ${\bf R}_i - {\bf a}_1$
and so on. The microscopic parameters for BaNi$_2$(PO$_4$)$_2$ ($S=1$) were
determined from the magnon dispersion as $J_1=0.38$~meV,  $J_3=1.52$~meV, and
$D=0.34$~meV \cite{Regnaultsup}. The second-neighbor exchange was estimated to be
much smaller $J_2=0.05$~meV and is neglected in the following.

Applying the Holstein-Primakoff transformation for two antiferromagnetic
sublattices and performing the Fourier transformation
\begin{equation}
a_i = \frac{1}{N^{1/2}} \sum_{\bf k} e^{i{\bf k}{\bf R}_i} a_{\bf k}\,, \quad
b_i = \frac{1}{N^{1/2}} \sum_{\bf k} e^{i{\bf k}({\bf R}_i+\bm{\rho})} b_{\bf k}\,,
\end{equation}
we obtain the harmonic part of the boson Hamiltonian
\begin{eqnarray}
\hat{\cal H}_2 & = & S \sum_{\bf k} \Bigl[ (3J_{13} +D)(a^\dagger_{\bf k} a^{_{}}_{\bf k}
+ b^\dagger_{\bf k} b^{_{}}_{\bf k})
\label{suppl:H2} \\
& - &  a_{\bf k} b_{-\bf k} F^*_{\bf k}
+  \frac{D}{2}\, (a_{\bf k} a_{-\bf k} + b_{\bf k} b_{-\bf k})+ \textrm{h.\,c.} \Bigr] \ ,
\nonumber
\end{eqnarray}
where we use the shorthand notations
\begin{eqnarray}
&&J_{13} = J_1 + J_3\, ,\\
&&F_{\bf k} =  J_1 \bigl(e^{i{\bf k}_1} +e^{i{\bf k}_2} + e^{i{\bf k}_3}\bigr)
+ J_3
\bigl(e^{i{\bf k}_4} +e^{i{\bf k}_5} + e^{i{\bf k}_6}\bigr),\nonumber
\end{eqnarray}
with ${\bf k}_n = {\bf k}\cdot {\bf r}_n$ and
\begin{eqnarray}
&& {\bf r}_1 = \frac{a}{\sqrt{3}}(1,0)\,, \quad 
{\bf r}_{2,3} =\frac{a}{\sqrt{3}}\Bigl(-\frac{1}{2},\pm\frac{\sqrt{3}}{2}\Bigr)\,,
\nonumber \\
&& {\bf r}_{4,5} = \frac{a}{\sqrt{3}} (1,\pm \sqrt{3})\ ,\quad
{\bf r}_6 =\frac{a}{\sqrt{3}}(-2,0) \ .
\end{eqnarray}

Diagonalization of the quadratic form (\ref{suppl:H2}) with the help of
the canonical Bogolyubov transformation yields
\begin{equation}
\hat{\cal H}_2 = \sum_{\bf k} \Bigl[ \varepsilon_\alpha({\bf k})
\alpha^\dagger_{\bf k} \alpha^{_{}}_{\bf k} + \varepsilon_\beta({\bf k})
\beta^\dagger_{\bf k} \beta^{_{}}_{\bf k} \Bigr] \ ,
\end{equation}
where excitation energies are
\begin{eqnarray}
&& \varepsilon_\alpha({\bf k}) =  S \sqrt{ (3J_{13} - |F_{\bf k}|)
(3J_{13} + |F_{\bf k}|+2D) }\ ,
\\[0.75mm]
&& \varepsilon_\beta({\bf k}) =  S \sqrt{ (3J_{13} + |F_{\bf k}|)
(3J_{13} - |F_{\bf k}| + 2D) }\ .
\end{eqnarray}
The first magnon branch is gapless, $\varepsilon_\alpha(0) = 0$,
and reaches the maximum value of
\begin{equation}
\omega^\alpha_{\rm max} = S \sqrt{(2J_3\!+4J_1)(4J_3\!+2J_1+2D)}
\approx 5.9\ \textrm{meV}
\label{suppl:Omax}
\end{equation}
at ${\bf k} = [q,q]$ with $q\approx 0.25$
in the reciprocal lattice units.
The second branch describes optical magnons with a finite energy gap
at ${\bf k}=0$
\begin{equation}
\Delta = 2S \sqrt{3DJ_{13}} \approx 2.8\ \textrm{meV} \ .
\end{equation}
The maximum of the optical branch $\omega^\beta_{\rm max}$ is close
to (\ref{suppl:Omax}).

\section*{Long-wavelength limit}

In the long-wavelength limit $k\to 0$ the energy of the acoustic branch has linear
dispersion $\varepsilon_\alpha({\bf k}) \approx c(ka)$ with the spin-wave velocity
\begin{equation}
c = S \sqrt{2 (J_{3} + {\textstyle \frac{1}{4}}J_1) (3J_{13}+D)}=4.42\ \textrm{meV} \ .
\end{equation}
For the optical branch one finds
\begin{equation}
\varepsilon_\beta({\bf k}) \approx \Delta + \frac{(ka)^2}{2m}\,, \ \
m = \frac{(3DJ_{13})^{1/2}}{S(J_3\! + {\textstyle \frac{1}{4}}J_1)(3J_{13}\!-D)}\,,
\end{equation}
with $m=0.16$~meV$^{-1}$ for BaNi$_2$(PO$_4$)$_2$.

For small $k$ the Bogolyubov transformation can be written explicitly in the
following way. First, we transform from the original Holstein-Primakoff bosons
$a_i$ and $b_i$ to their linear combinations:
\begin{equation}
\bar{a}_i = \frac{1}{\sqrt{2}}(a_i-b_i) \ , \quad
\bar{b}_i = \frac{1}{\sqrt{2}}(a_i+b_i)  \ .
\end{equation}
The Fourier transformed Hamiltonian (\ref{suppl:H2}) takes the following
form
\begin{eqnarray}
\hat{\cal H}_2 & = & S \sum_{\bf k} \Bigl[ (3J_{13} +D)(\bar{a}^\dagger_{\bf k} \bar{a}^{_{}}_{\bf k}
+ \bar{b}^\dagger_{\bf k} \bar{b}^{_{}}_{\bf k})
\label{suppl:H2ab} \\
& + &  
\frac{1}{2}(D+|F_{\bf k}|)\,\bar{a}_{\bf k} \bar{a}_{-\bf k}\! +
\frac{1}{2}(D-|F_{\bf k}|)\,\bar{b}_{\bf k} \bar{b}_{-\bf k}\! + \textrm{h.\,c.} \Bigr] \,.
\nonumber
\end{eqnarray}
Second, the standard $u$--$v$ transformation is applied separately for $\bar{a}_{\bf k}$
and $\bar{b}_{\bf k}$ bosons. In particular, for the acoustic branch,
$\bar{a}_{\bf k}=u_k\alpha_{\bf k} + v_k \alpha^\dagger_{-\bf k}$,
we obtain
\begin{eqnarray}
u_k & \approx & \sqrt{\frac{1+d}{2\tilde{c}ka}} \Bigl(1+ \frac{\tilde{c}ka}{2(1+d)}\Bigr) \ ,
\nonumber \\
v_k & \approx & -\sqrt{\frac{1+d}{2\tilde{c}ka}} \Bigl(1- \frac{\tilde{c}ka}{2(1+d)}\Bigr)  \ ,
\label{suppl:uva}
\end{eqnarray}
where  $d=D/(3J_{13})$ and $\tilde{c} = c/(3J_{13})$.
In the case of  BaNi$_2$(PO$_4$)$_2$ the two dimensionless constants are
$d\approx 0.06$ and $\tilde{c} \approx 0.77$.

Similarly, for optical magnons with $k \to 0$ we obtain
$\bar{b}_{\bf k} = u_{\bf k}\beta_{\bf k} + v_{\bf k}\beta^\dagger_{-\bf k}$ with
\begin{equation}
u_0^2 + v_0^2 = \frac{1+d}{2\sqrt{d}}\ , \quad 2u_0v_0 =  \frac{1-d}{2\sqrt{d}} \ .
\end{equation}

\section*{Magnon-magnon interaction}

For a collinear antiferromagnet the interaction between spin-waves  is described
by four-magnon terms in the bosonic Hamiltonian. The four-magnon terms of the
exchange origin are expressed as
\begin{eqnarray}
H^{\rm (ex)}_4 & = &-\frac{1}{N} \sum_{1+2=3+4}F_{{\bf k}_3-{\bf k}_2}
a_4^\dagger b_3^\dagger b_2 a_1
\label{HJ}
\\
& + &\frac{1}{4N} \sum_{1+2+3=4}F_{{\bf k}_1}\Bigl(
a_4^\dagger a_3 a_2 b_1 + b_4^\dagger b_3 b_2 a_1 + \textrm{h.\,c.}\Bigr),
\nonumber
\end{eqnarray}
where $a_1$ stands for $a_{{\bf k}_1}$ etc. The single-ion anisotropy contributes
\begin{eqnarray}
H^{\rm (an)}_4 & = &-\frac{D}{2N} \sum_{1+2=3+4}\bigl(
a_4^\dagger a_3^\dagger a_2 a_1 + b_4^\dagger b_3^\dagger b_2 b_1\bigr)
\label{HD}
\\
& - &\frac{D}{4N} \sum_{1+2+3=4}\bigl(
a_4^\dagger a_3 a_2 a_1 + b_4^\dagger b_3 b_2 b_1 + \textrm{h.\,c.}\bigr).
\nonumber
\end{eqnarray}

Performing    transformation from $a_{\bf k}$, $b_{\bf k}$ to $\alpha_{\bf k}$,
$\beta_{\bf k}$ we obtain various magnon-magnon terms.
The scattering of optical ($\beta$) magnons
on each other, which will be referred to as the roton-roton interaction,
can be straightforwardly written as
\begin{eqnarray}
V^{\rm rr} & = &-\frac{3J_{13}+D}{N} \sum_{1+2=3+4}
\beta_4^\dagger \beta_3^\dagger \beta_2 \beta_1 \ .
\end{eqnarray}
Derivation of the  roton-phonon interaction (scattering of the optical magnon on
the acoustic one, $\beta$ on $\alpha$) is more involved and we obtain
an estimate as
\begin{equation}
V^{\rm rp}  = -\frac{3(J_3 + \frac{1}{4}J_1)(1+d)^2}{4\sqrt{3d}\,
\tilde{c}N}\!\!\!\sum_{1+2=3+4}\!\!\!\sqrt{k_2k_3}\,\beta_4^\dagger
\alpha_3^\dagger \alpha_2 \beta_1 \ .
\label{Vrp}
\end{equation}

The individual terms in the magnon-magnon interaction obtained from (\ref{HJ}) and (\ref{HD}) 
applying the Bogolyubov transformation are proportional to $u_{\bf q}u_{\bf q'}$ 
and diverge for scattering processes involving acoustic magnons, see (\ref{suppl:uva}). However, 
the leading  $\sim {\cal O}(1/\sqrt{qq'})$ and the subleading singularity $\sim {\cal O}(1)$  cancel out
in their net contribution and $V^{\rm rp}_{\bf kq;k'q'}\propto \sqrt{qq'}$
in agreement with the hydrodynamic approach \cite{Baryakhtar73sup,LLIXsup}.

Local modulation of magnetic coupling constants due to structural disorder,  etc., will result in {\it independent} variations
of $J$- and $D$-terms in magnon-magnon interaction in (\ref{HJ}) and (\ref{HD}).
Thus, the resultant impurity-assisted magnon-magnon interaction will retain the same structure as the
magnon-magnon interaction, with two important differences. First, the momentum in such a scattering is not conserved,
and, second, the variation of $J$ ($\delta J$) is associated only with (\ref{HJ}) and the  variation
$\delta D$ will contain only (\ref{HD}) part. Since such variations are independent,
it suffices to consider one of them and treat the associated constant as a free parameter.
The most important consequence of this consideration is that,
in the impurity scattering, there is
no cancellation of the individual terms that are proportional to
$u_{\bf q}u_{\bf q'}\propto 1/\sqrt{qq'}$,  compared to the case of magnon-magnon scattering in (\ref{Vrp})
discussed above where such a cancellation does take place. Thus, in the long-wavelength limit,
$V^{\rm\, imp}_{\bf kq;k'q'}\propto 1/\sqrt{qq'}$, with a coefficient proportional to the impurity concentration and
strength of the disorder.

\section*{Relaxation rate of optical magnons}

The lowest-order diagram for the magnon self-energy
calculated using  Matsubara technique is
\begin{eqnarray}
\Sigma({\bf k},i\omega)  & = & \frac{1}{2}\sum_{{\bf q},{\bf q}'}
\frac{\big|V_{\bf k}({\bf q},{\bf q}')\big|^2}
{i\omega + \varepsilon_{\bf q}\! - \varepsilon_{\bf q'}\!-\varepsilon_{\bf k'}}
\\
& \times & [n_{\bf q}(n_{\bf q'}\!+1)(n_{\bf k'}+1) -(n_{\bf q}\!+1)n_{\bf q'}n_{\bf k'}]
\nonumber
\end{eqnarray}
with $\bf k' = k+q-q'$. Then the damping rate is
\begin{eqnarray}
\Gamma_{\bf k} &=&
 \frac{\pi}{2}\,\sum_{{\bf q},{\bf q}'}
\big|V_{\bf k}({\bf q},{\bf q}')\big|^2 [n_{\bf q}(1\!+n_{\bf q'}\!+n_{\bf k'}) -n_{\bf q'}n_{\bf k'}]
\nonumber\\
&\times&\delta(\varepsilon_{\bf k} + \varepsilon_{\bf q}
-\varepsilon_{\bf q'} - \varepsilon_{\bf k'})\,.
\label{GammaK}
\end{eqnarray}

First, we consider the roton-roton scattering processes. The low-temperature asymptote of
(\ref{GammaK}) in this case is obtained by taking $T\ll\Delta$ and keeping the leading exponentially small
occupation factor. Then, for an optical magnon with ${\bf k}=0$ in two dimensions
\begin{eqnarray}
\Gamma_0^{\rm rr} &=&  \frac{3V_{\rm rr}^2}{64\pi^2}\int_0^\infty  q\,dq
\int_0^q q'\, dq'\int_0^{2\pi} d\varphi \; n_{\bf q}
\label{Gamma01}\\
&&\times\ \delta\left(\frac{q^2}{2m}
-\frac{q'^2}{2m}-\frac{|{\bf q}-{\bf q'}|^2}{2m}\right) .\nonumber
\end{eqnarray}

Performing integration in (\ref{Gamma01}) and using parameters for  BaNi$_2$(PO$_4$)$_2$ discussed above we obtain
\begin{equation}
\Gamma_0^{\rm rr} =  \frac{3(mV_{\rm rr})^2T}{64\pi}\, e^{-\Delta/T} \approx 0.035\,\frac{T}{\Delta}\,e^{-\Delta/T}
\textrm{[meV]} \ .
\label{Gamma02rr}
\end{equation}
Without going into details, there exist another channel of scattering that corresponds to a conversion
of two rotons into two high-energy phonons,
$\beta\beta\rightarrow\alpha\alpha$, which leads to the decay rate of the optical mode of the same exponential form
as in (\ref{Gamma02rr}) with a numerical coefficient of the same order.

Finally, the the low-temperature asymptote of the roton-phonon scattering in
(\ref{GammaK}) is
\begin{eqnarray}
\Gamma_0^{\rm rp} & = & \frac{3\tilde{V}_{\rm rp}^2}{32\pi^2}\int_0^\infty\! q^2
\,dq\int_0^q q'^2\, dq'\int_0^{2\pi} \!d\varphi \: n_{\bf q}(n_{\bf q'}\!+1)
\nonumber \\
& \times &\delta\left(cq-cq'
-\frac{|{\bf q}-{\bf q'}|^2}{2m}\right) ,
\label{Gamma03}
\end{eqnarray}
where $V_{\rm rp} = \tilde{V}_{\rm rp}\sqrt{qq'}$. Subsequent integration yields
\begin{equation}
\Gamma_0^{\rm rp} =  \frac{\pi^3}{20}\frac{\tilde{V}_{\rm rp}^2T^5}{c^6} \approx 0.18\,\biggl(\frac{T}{\Delta}\biggr)^5
\textrm{[meV]} \ .
\label{Gamma02}
\end{equation}


\end{document}